\documentclass[%
 amsmath,amssymb,
 aps, physrev,
 pre,
 twocolumn,
]{revtex4-2}

\usepackage{graphicx}
\usepackage{dcolumn}
\usepackage{bm}

\usepackage{color}
\usepackage{amssymb}
\usepackage{natbib}
\usepackage{hyperref}
\usepackage{xcolor}
\usepackage{float}

\begin{document}

\preprint{APS/123-QED}

\title{
A novel scheme for measuring the growth of Alfv\'en wave parametric decay instability using counter-propagating waves
}

\author{Feiyu Li$^{1}$}
    \altaffiliation[Contact author: lif02501@gmail.com]{}
\author{Seth Dorfman$^{2,3}$}
\author{Xiangrong Fu$^{1,4}$}
\affiliation{
$^{1}$ New Mexico Consortium, Los Alamos, New Mexico 87544, USA\\
$^{2}$ Space Science Institute, Boulder, Colorado 80301, USA\\
$^{3}$ University of California Los Angeles, Los Angeles, California 90095, USA\\
$^{4}$ Los Alamos National Laboratory, Los Alamos, New Mexico 87545, USA
}
    
\date{\today}

\begin{abstract}

The parametric decay instability (PDI) of Alfv\'en waves--where a pump Alfv\'en wave decays into a backward-propagating child Alfv\'en wave and a forward ion acoustic wave---is a fundamental nonlinear wave-wave interaction and holds significant implications for space and laboratory plasmas. However, to date there has been no direct experimental measurement of PDI. 
Here, we propose a novel and experimentally viable scheme to quantify the growth of Alfv\'en wave PDI on a linear device using a large pump Alfv\'en wave and a small counter-propagating seed Alfv\'en wave, with the seed wave frequency tuned to match the backward Alfv\'en wave generated by standard PDI. 
Using hybrid simulations, we show that energy transfer from the pump to the seed reduces the latter’s spatial damping. By comparing seed wave amplitudes with and without the pump wave, this damping reduction can be used as a direct and reliable proxy for PDI growth. The method is validated in our simulations across a range of plasma and wave parameters and agrees well with theoretical predictions. Notably, the scheme exhibits no threshold for PDI excitation and is, in principle, readily implementable under current laboratory conditions.
This scheme is a critical step toward solving the challenge of experimentally accessing Alfv\'en wave PDI and provides an elegant method that may be used to validate fundamental theories of parametric instabilities in controlled laboratory settings.  

\end{abstract} 

\maketitle

\section{Introduction}

\begin{figure*}[htp]
	\centering
	\includegraphics[width=0.9\textwidth]{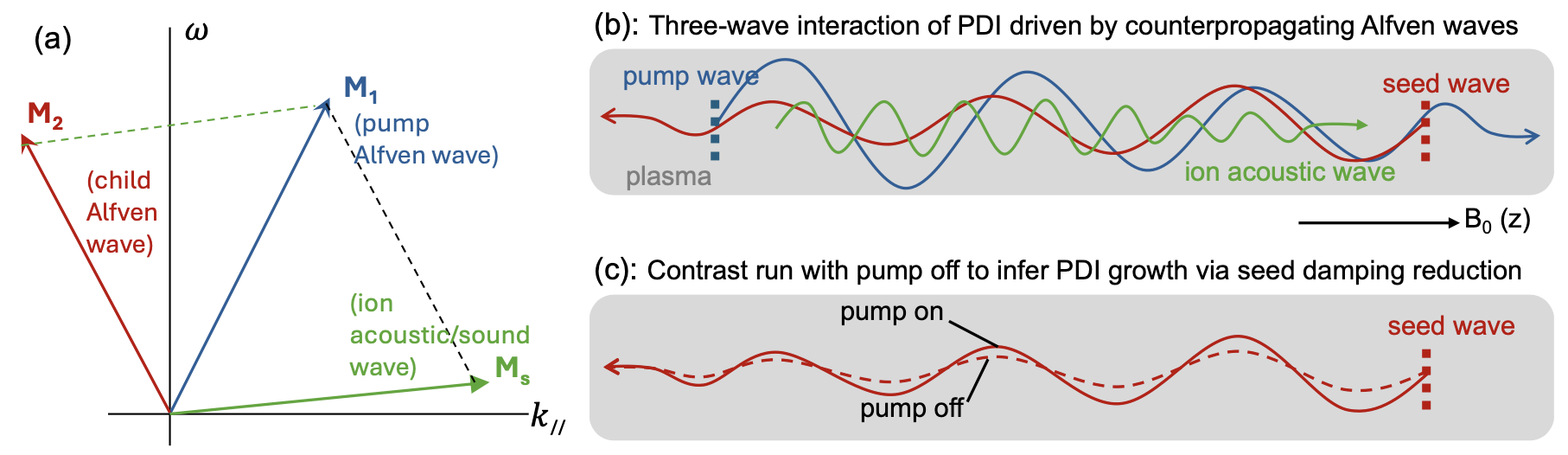}
	\caption{(a) $\omega$-$k_\parallel$ diagram of PDI. (b, c) A sketch of the scheme to measure PDI growth using two counter-propagating Alfv\'en waves. In (b) a beat acoustic mode is expected and in (c) the pump is turned off and the seed wave damping reduction with the pump on vs off is used to infer PDI growth.}
	\label{fig1}
\end{figure*}

Parametric instabilities play a pivotal role across a variety of physical systems, shaping the dynamics of energy transfer and nonlinear mode coupling. In laser-driven inertial confinement fusion, critical parametric processes such as stimulated Raman/Brillouin scattering directly impact energy deposition and particle acceleration~\cite{hurricane2023physics,montgomery2016two,klimo2013laser}. 
In magnetically confined fusion plasmas, parametric instabilities can influence confinement properties and the efficiency of heating schemes through the redistribution of wave energy~\cite{hasegawa1976parametric,sharma2021kinetic,clod2024cascades}. 
Parametric processes also underpin a range of phenomena in nonlinear optics~\cite{chirkin2000consecutive,wright2022physics,mckenna2022ultra,ledezma2022intense}, and more broadly in nonlinear science, where they serve as prototypical mechanisms for the onset of complex behavior~\cite{kadri2013generation,May_Qin_2024}. 
The present work focuses on the parametric decay instability (PDI) of Alfv\'en waves, 
which are a fundamental mode in magnetized plasmas and ubiquitously observed in space~\cite{belcher1971large,knudsen1992alfven,de2007chromospheric,kasper2019alfvenic,lysak2023kinetic} and laboratory~\cite{chen2016physics,gekelman2011many,carter2006laboratory,howes2012toward,schroeder2021laboratory}. In PDI a large pump Alfv\'en wave (mode $M_1)$ resonantly couples to a backward-propagating child Alfv\'en wave ($M_2$) and a forward-propagating ion acoustic/sound wave ($M_s)$, 
satisfying the frequency and wavevector matching conditions $\omega_1=\omega_2+\omega_s$ and $\vec{k}_{1}=\vec{k}_{2}+\vec{k}_{s}$ [Fig.~\ref{fig1}(a)].
Alfv\'en wave PDI directly drives cascade decays and influences the onset and dissipation of plasma turbulence~\cite{del2001parametric,howes2013alfven,chandran2018parametric}, making it a key nonlinear process in both laboratory~\cite{hasegawa1976parametric,qiu2018nonlinear} and astrophysical plasmas~\cite{malara2022parametric}. In particular, PDI is widely believed to contribute to solar coronal heating~\cite{shoda2018frequency} and solar wind acceleration~\cite{Rivera2024}. Observational evidence of PDI has been reported in near-Sun atmosphere~\cite{hahn2022evidence} and in solar wind at 1 AU~\cite{bowen2018density}. 

Laboratory experiments offer the unique ability to isolate, control and validate individual plasma processes.
Yet, direct laboratory observation of Alfv\'en wave PDI remains elusive, despite decades of theoretical~\cite{derby1978modulational,goldstein1978instability,hollweg1993modulational} and numerical~\cite{del2001parametric,fu2018parametric,li2022parametric,nariyuki2008parametric} studies.   
Related processes have been reported in both linear devices~\cite{dorfman2013nonlinear,dorfman2016observation} and fusion tokamaks~\cite{zhu2022nonlinear}. 
With the Large Plasma Device (LAPD)~\cite{gekelman2016upgraded},
the three-wave interaction of PDI was demonstrated by driving an ion acoustic beat wave using two counter-propagating Alfv\'en waves of \emph{comparable} amplitudes~\cite{dorfman2013nonlinear}. Similarly, on the Experimental Advanced Superconducting Tokamak, a geodesic beat acoustic mode was excited by multiple Alfv\'en eigenmodes~\cite{zhu2022nonlinear}. However, no instability behavior---i.e. exponential growth characteristic of PDI---was observed in those experiments.

Exciting PDI with a single pump Alfv\'en wave typically requires the instability growth rate to exceed the geometric mean of the damping rates of the two child modes ($\Gamma_2$, $\Gamma_s$)~\cite{Nishikawa1968,li2024}, 
\begin{equation}
\label{gamma_g_th}
    \gamma_g/\omega_1=\frac{1}{2}\frac{\delta B_1}{B_0}/\beta^{1/4} >\sqrt{\Gamma_2\Gamma_s}/\omega_1,
\end{equation}
where $\gamma_g$
is the ideal (undamped) growth rate~\cite{sagdeev1969nonlinear,Nishikawa1968}, with $\delta B_1$ the pump wave magnetic field amplitude, $B_0$ the background magnetic field, and $\beta$ the plasma beta.
This criterion highlights the strong damping sensitivity of PDI in laboratory conditions. 
Recent studies suggest that satisfying Eq.~(\ref{gamma_g_th}) remains difficult under current experimental setups both for LAPD-like linear devices~\cite{li2024} and fusion tokamaks~\cite{qiu2018nonlinear}. 

\section{Proposed scheme for PDI growth measurement}
In this work, we propose a novel and experimentally viable scheme for measuring the growth of Alfv\'en wave PDI on a linear device and demonstrate our method using hybrid simulations. 
To circumvent the threshold constraint described by Eq.~(\ref{gamma_g_th}), a key innovation involves launching a \emph{small} counter-propagating Alfv\'en wave, a few wavelengths downstream from the pump. The seed wave acting as the child Alfv\'en wave ($M_2$) seeds the decay process of the pump wave. 
A schematic of this interaction is shown in Fig.~\ref{fig1}(b), where the beat interaction between the pump and seed generates a sound wave ($M_s$). The seed wave is continuously driven to achieve a steady-state condition with constant $M_2$ amplitude at any fixed location; this corresponds to no temporal damping (i.e. $\Gamma_2=0$), reducing the threshold condition Eq.~(\ref{gamma_g_th}) to $\gamma_g>0$.
In other words, PDI can now be excited without any intrinsic threshold.

While the continuously driven seed wave has no temporal damping, spatial damping of Alfv\'en waves due to a combination of electron Landau damping, collisional damping, and geometric attenuation remains significant under typical laboratory conditions~\cite{morales1997structure,thuecks2009tests,bose2019measured,li2024}.
The proposed configuration is conceptually similar to Raman amplification in laser-plasma instabilities, where a long pump laser amplifies a short, counter-propagating seed pulse to ultra-high intensities~\cite{trines2011simulations}. 
Likewise, the pump Alfv\'en wave in our scheme is expected to transfer energy to the seed wave, offsetting the spatial damping of the latter. This ``damping reduction'' (or growth/amplification) of the seed wave---measured by comparing its amplitude with the pump on vs off, as sketched in Fig.~\ref{fig1}(c)---serves as a diagnostic for PDI growth.

When the damping rates of the child modes are non-negligible, the effective PDI growth rate deviates from the ideal value $\gamma_g$ [see Eq.~(\ref{gamma_g_th})]. In this more general case, the growth rate is given by $\gamma_g^{\rm eff}=\frac{1}{2}\big[-(\Gamma_2+\Gamma_s)+\sqrt{(\Gamma_2-\Gamma_s)^2+4\gamma_g^2}\big]$~\cite{Nishikawa1968}. 
Substituting $\Gamma_2=0$, as appropriate for our seeded configuration, the effective growth rate simplifies to:
\begin{equation}
\label{gamma_g_eff2}
    \gamma_g^{\rm eff}=\frac{1}{2}\big(-\Gamma_s+\sqrt{\Gamma_s^2+4\gamma_g^2}\big). 
\end{equation}
The fact that $\gamma_g^{\rm eff}$ in Eq.~(\ref{gamma_g_eff2}) is always positive reaffirms the threshold-less excitation for our scheme.
The objective of this work is to quantitatively measure the growth of the seed wave in hybrid simulations and compare it with theoretical prediction from Eq.~(\ref{gamma_g_eff2}). 

\begin{figure*}[htp]
	\centering
	\includegraphics[width=0.96\textwidth]{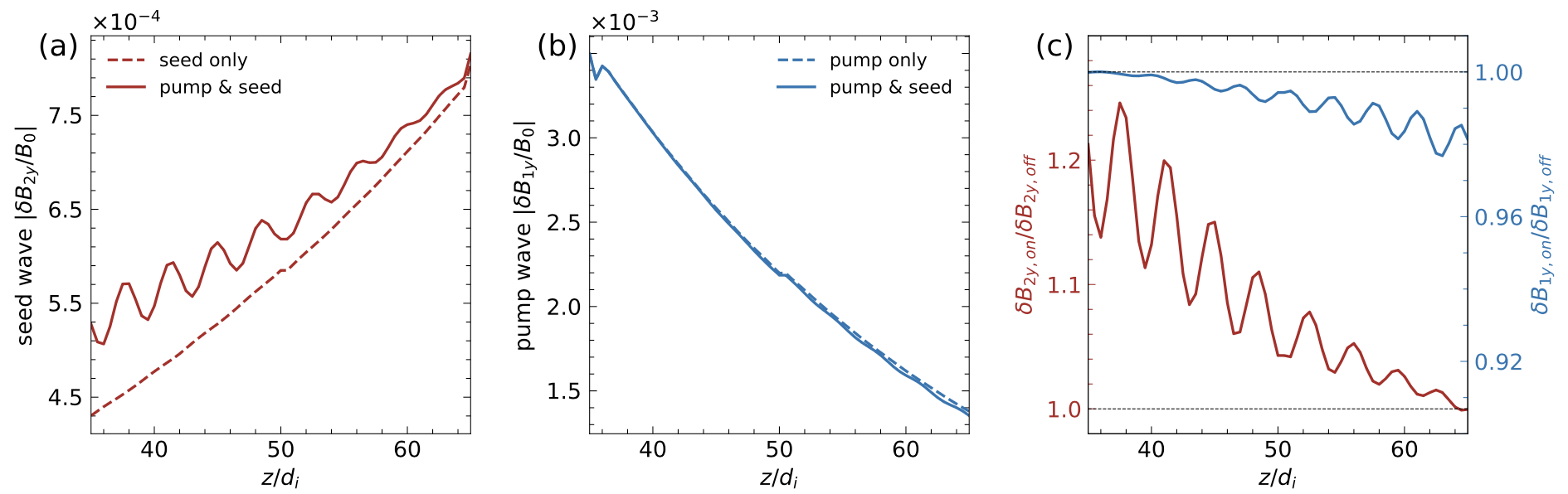}
	\caption{A 1D hybrid simulation example. (a) Seed wave amplitude vs $z$ in pump-off (dashed) and pump-on runs (solid). (b) Pump wave amplitude vs $z$ in seed-off (dashed) and seed-on runs (solid). (c) Corresponding seed amplitude ratios between pump on and off (red) and pump amplitude ratios between seed on and off (blue). }
	\label{fig2}
\end{figure*}

\section{Results}

\subsection{Simulation setup and a selected case}

We begin with a one-dimensional (1D) simulation using the hybrid (kinetic ions and a massless electron fluid) code H3D~\cite{karimabadi2006global}, which was recently adapted to study single-wave PDI in configurations relevant to the LAPD~\cite{li2022hybrid,li2024}. 
In our setup, a uniform hydrogen plasma occupies $z\in$ [0, 100] $d_i$ under a constant background magnetic field, where $d_i$ is the ion inertial length. The plasma beta is $\beta=1.53\times 10^{-3}$, and the electron-to-ion temperature ratio $T_e/T_i=5.65$.
The electron fluid follows the adiabatic equation of state, $T_e/n_e^{\gamma_e-1}=\rm const$, with $\gamma_e=5/3$.  
The spatial resolution is $\Delta z=0.5d_i$, with each cell initialized with 10000 ions. 
The time step is $\Delta t=0.01 \Omega_{ci}^{-1}$, where $\Omega_{ci}$ is the ion cyclotron frequency.

A pump Alfv\'en wave is injected at $z=35 d_i$, and a counter-propagating seed wave is injected at $z=65 d_i$.
The normalized amplitudes at injection are $\delta B_1/B_0=4\times 10^{-3}$ (pump) and $\delta B_2/B_0=1\times 10^{-3}$ (seed). Both waves are left-hand circularly polarized.
The normalized frequencies are $\omega_1/\Omega_{ci}=0.63$ and $\omega_2=0.91\omega_1$, such that the frequency difference $\omega_1-\omega_2\simeq \omega_s\simeq 2\sqrt{\beta}\omega_1$, satisfying resonance conditions for driving a beat ion acoustic mode~\cite{dorfman2013nonlinear}. 
We adopt finite-frequency waves to mirror the experimental need to fit several wave cycles into the device~\cite{gigliotti2009generation,dorfman2016observation} and check that the associated two-fluid effects~\cite{hollweg1994beat} on the PDI growth rate [Eq.~(\ref{gamma_g_th})] are negligible in the low-beta regime. 
Wave injection is achieved by prescribing the electric field at each wave's respective injection location. The wave amplitude ramps up over the first $50\Omega_{ci}^{-1}$ and then remains constant until the end of the simulation at $t_{\rm max}=2000\Omega_{ci}^{-1}$. 
The Alfv\'en waves are absorbed on both sides, $z\in$ [0, 30] $d_i$ and $z\in$ [70, 100] $d_i$, using field masks~\cite{li2022parametric}.
Under periodic boundary conditions for particles the circulating ions do not interfere with the central region 
$z\in$ [35, 65] $d_i$ within the timescale of interest, due to 
the wide absorption zones. 
To isolate the PDI growth, contrast runs with the pump turned off are also performed, following the schematic of Fig.~\ref{fig1}(c).  
Notice that the interaction of counter-propagating waves in our scheme does not lead to Alfv\'enic turbulence. This is because: i) The relevant nonlinear interaction generally requires that the two waves have perpendicular wave vectors with a nonzero cross product, i.e. $k_{\perp,1}\times k_{\perp,2}\neq 0$~\cite{howes2012toward}, and $k_\perp=0$ for our 1D simulations; ii) even with finite $k_\perp$ under a more realistic 3D configuration, the Alfv\'enic modes produced by the nonlinear interaction of counter-propagating Alfven waves will be nonresonant quasi-modes, and the process is expected to be much less effective than the resonant coupling to the natural acoustic mode in the parallel direction which underlies the PDI.

Figure~\ref{fig2}(a) shows the spatial profile of the seed wave, measured in simulations with the pump off (dashed curve) and on (solid curve). 
The amplitude at each $z$ represents the spectral component of the magnetic field (specifically $B_y$) at frequency $\omega=\omega_2$, probed over the time window $t\in$ [500,1500] $\Omega_{ci}^{-1}$.
As expected, the seed wave---launched at $z=65\ d_i$---experiences spatial damping as it propagates leftward. However, this damping is noticeably reduced when the pump is present, consistent with energy transfer via PDI.
To quantify this effect, the red curve in Fig.~\ref{fig2}(c) shows the amplitude ratio $\delta B_{2y,\rm on}/\delta B_{2y,\rm off}$, which remains consistently above unity and increases toward smaller $z$. By the time the seed wave reaches the pump injection location ($z=35 d_i$), its amplitude is enhanced by approximately 20\% compared to the pump-off baseline. 

To trace the source of this amplification, we performed a complementary contrast run with only the pump wave present, and compared the pump wave amplitude under conditions with and without the seed. Figure~\ref{fig2}(b) displays the pump wave profiles in both cases. The difference is subtle, but better visualized in Fig.~\ref{fig2}(c), where the blue curve shows the ratio $\delta B_{1y,\rm on}/\delta B_{1y,\rm off}$. This ratio remains below unity, indicating that the pump wave loses energy in the presence of the seed. The relative amplitude decrease reaches $\sim$2\% at $z=65 d_i$, where the seed is injected. Notably, the percentage gain in the seed wave is larger than the percentage loss in the pump. This asymmetry reflects the difference in energy content between the two waves---the pump carries significantly more energy, and a small fractional loss is sufficient to cause a noticeable amplification in the seed.

While the preceding contrast runs demonstrate the potential to measure PDI growth via seed wave difference, we also observe an unexpected modulation in the spatial profiles of the wave amplitude ratios. 
This modulation was traced to few-percent-level reflection of the seed (pump) wave at the pump (seed) injection location, likely due to local field discontinuities~\cite{li2024}. 
Physically, reflections combine with the incident waves to form a partial standing wave, which manifests as spatial oscillations in the wave amplitude---and consequently, spatial oscillations in the amplitude ratio. 
To remove the influence of reflection and isolate PDI-induced amplification, we apply an envelope-averaging technique: extract and average the top and bottom envelopes of the modulated amplitude ratio curves. 
A simple model justifying this approach is provided in Appendix~\ref{sec:reflection}. All amplitude ratio data presented in the remainder of the paper incorporate this envelope-averaging correction.

\begin{figure*}[htp]
	\centering
	\includegraphics[width=0.96\textwidth]{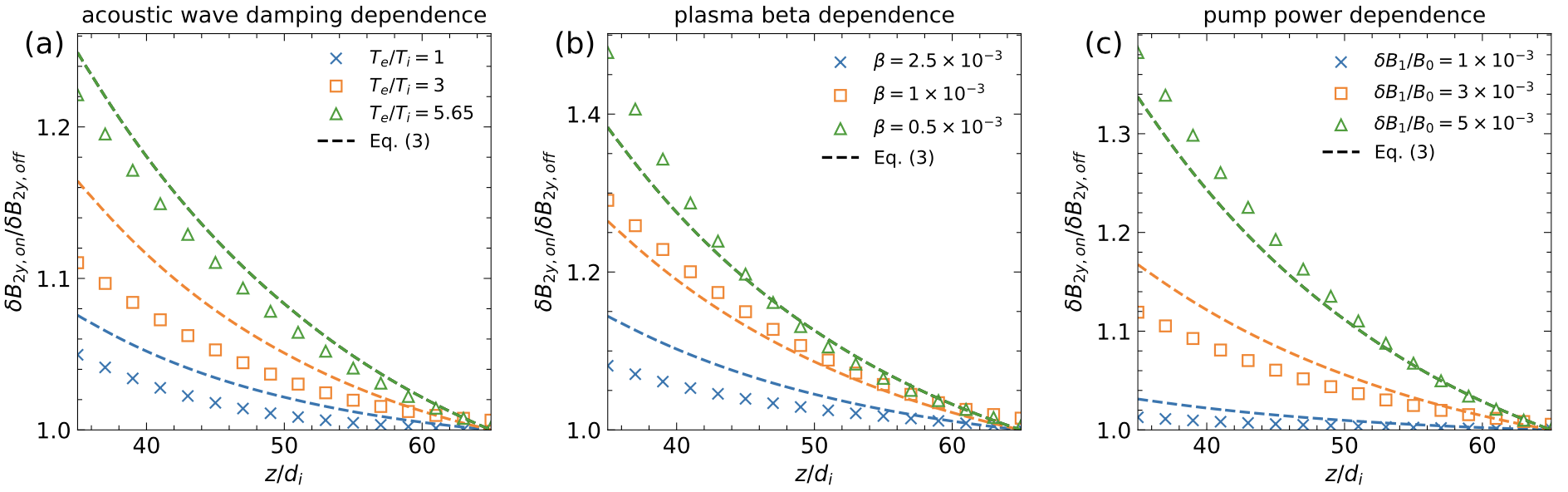}
	\caption{Comparison between measured PDI growth (scatters) and Eq.~(\ref{wave_ratio}) (dashed curves). (a) Dependence of seed amplitude ratios (vs $z$) on the acoustic wave damping, through varying the electron-to-ion temperature ratio. (b) Plasma beta dependence. (c) Pump power/amplitude dependence. See the main text for the detailed parameters used.}
	\label{fig3}
\end{figure*}

\subsection{A model for convective PDI growth and its comparison with simulations}

To compare the measured seed wave amplification (i.e. the amplitude ratios) with theoretical expectations, we first construct a spatial profile of the anticipated growth using Eq.~(\ref{gamma_g_eff2}). 
Let us discretize the domain as $n$ uniform bins of size $\delta z$ with edges at $[z_0, z_1, ..., z_n]$, 
where the pump is injected at $z_0$ and the seed at $z_n$. 
The seed wave amplitude will decrease due to spatial damping as the wave traverses a single bin from $z_i$ to $z_{i-1}$; with the pump off, this change can be modeled as $\delta B_{2,\rm off}(z_{i-1})=\delta B_{2,\rm off}(z_i)\exp(-\gamma_d\delta t)$.  Here, $\delta B_{2,\rm off}(z)$ is the seed amplitude with the pump off, $\delta t=\delta z/v_g$ (where $v_g$ is the seed wave group velocity) is the time required for a seed wavepacket to traverse a single bin, and $\gamma_d/v_g$ is the spatial damping rate.  When the pump is on, the seed wave amplitude will also grow over the time interval $\delta t$ due to PDI; as a result, after traversing only a single bin, the seed wave will be larger in amplitude by a factor of $\exp[\gamma_g^{\rm eff}(z_i)\delta t]$ when the pump is on than it is when the pump is off.  Note that the PDI growth rate $\gamma_g^{\rm eff}$ is a function of $z$ due to the spatial damping of the pump wave, which modulates the local pump amplitude $\delta B_1(z)$.  To determine the seed wave amplitude ratio $R(z_i)\equiv\delta B_{2,\rm on}(z_i)/\delta B_{2,\rm off}(z_i)$, we multiply together the factors by which the seed wave grows or damps in each bin it traverses, while noting that at the seed wave injection location $\delta B_{2,\rm on}(z_n)=\delta B_{2,\rm off}(z_n)$:
\begin{equation}
\label{wave_ratio}
\begin{split}
   R(z_i) &=\frac{\delta B_{2,\rm on}(z_n)\exp[\sum_{j>i} (\gamma_g^{\rm eff}(z_j)-\gamma_d)\delta t]}{\delta B_{2,\rm off}(z_n)\exp[\sum_{j>i} (-\gamma_d\delta t)]}\\
   &=\exp[\sum_{j>i} \gamma_g^{\rm eff}(z_j)\delta t].
\end{split}
\end{equation}
This expression depends solely on the effective growth rate $\gamma_g^{\rm eff}$, which is itself dependent on the spatially varying pump amplitude via Eq.~(\ref{gamma_g_eff2}).

We conducted a series of 1D simulations to compare the measured seed wave amplitude ratios with theoretical predictions from Eq.~(\ref{wave_ratio}). Since the effective growth rate $\gamma_g^{\rm eff}$ depends on the sound wave damping rate $\Gamma_s$, the plasma beta $\beta$, and the normalized pump amplitude $\delta B_1/B_0$ [see Eq.~(\ref{gamma_g_eff2})], we varied each of these parameters to isolate their effects. 
First, we varied the electron-to-ion temperature ratio $T_e/T_i\equiv \Theta$ to study how sound wave damping influences PDI, while keeping $\beta=1.53\times 10^{-3}$, $\delta B_1/B_0=4\times 10^{-3}$, and $\delta B_2/B_0=1\times 10^{-3}$ fixed. 
In these simulations, $\Gamma_s$ is dominated by ion Landau damping and we estimate it as $\Gamma_s/\omega_s=1.1\times \Theta^{-7/4}\exp(-\Theta^{-2})$ for $1<\Theta<10$~\cite{chen2015introduction}.  
Second, we varied plasma beta $\beta$, while holding $\Theta=4$, $\delta B_1/B_0=4\times 10^{-3}$, and $\delta B_2/B_0=1\times 10^{-3}$ constant. 
Third, we scanned pump wave amplitude $\delta B_1/B_0$, while fixing $\beta=1.53\times 10^{-3}$, $\Theta=5.65$, and $\delta B_2/B_0=1\times 10^{-3}$. 
For all simulations, the pump wave frequency was fixed at $\omega_1/\Omega_{ci}=0.63$ and the seed wave frequency $\omega_2$ was chosen based on $\beta$ such that the counter-propagating Alfv\'en waves resonantly drive a beat acoustic mode. 
Figure~\ref{fig3} summarizes the simulation results across all three categories. Overall, we observe good agreement with theory: weaker acoustic wave damping (bigger $\Theta$), smaller $\beta$, and higher pump amplitudes all lead to larger PDI growth. 
Minor deviations from the theoretical predictions likely stem from the approximate nature of the empirical ion Landau damping formula, as well as any plasma modifications (e.g. $\beta$ evolution) caused by the nonlinear interactions. In particular, in several cases, the excited acoustic wave exhibits strong nonlinearity (e.g. wave steepening, phase mixing, and/or ion trapping), and as the acoustic wave propagates, it may develop different strengths of nonlinearity at different locations; both phenomena may affect the local acoustic damping rate and are not captured by the linear, time-only model for the damping rate.

\section{Experimental feasibility}

The present study adopts parameters and configurations that closely mirror those achievable in the LAPD, making the proposed scheme particularly promising for experimental realization. Yet, unlike well-defined parameters in simulations, experimental plasma parameters are subject to uncertainty; 
thus, future experimental studies will use a seed wave frequency scan to confirm the resonance condition, following the approach taken in prior beat acoustic wave experiments~\cite{dorfman2013nonlinear}. 
Moreover, the wave reflection at the injection locations seen in our simulations is likely to have an analogue in experiments. In particular,
the high-current loop antennas used to launch nonlinear Alfv\'en waves on LAPD~\cite{gigliotti2009generation} may perturb the local magnetic field and plasma and cause reflections; reflection from a background magnetic field gradient has been observed experimentally~\cite{bose2024experimental}. 
If reflection happens in PDI measurement, tailored experimental diagnostics (e.g. careful choice of probe number and placement or even an array of closely positioned probes) may be required to map out the full profile of wave amplification. 
Finally, acoustic wave damping is expected to be stronger in experiments due to ion-neutral collisions, which are absent in the simulation model. This enhanced damping will reduce the effective PDI growth rate  following Eq.~(\ref{gamma_g_eff2}). 
Overall, to maximize the feasibility of experimental PDI measurements, conditions may be optimized to the extent possible to minimize acoustic wave damping
and reduce potential wave reflections.

\section{Summary}

In summary, we have used hybrid simulations---carefully designed to reflect realistic laboratory conditions---to demonstrate a novel and viable method for measuring the growth of Alfvén wave PDI. By launching counter-propagating pump and seed waves and implementing a tailored set of contrast runs, we introduced a robust scheme that effectively bypasses the threshold constraint inherent in traditional single-wave PDI excitation. 
Through a systematic exploration across a range of wave and plasma parameters, we observed good agreement between theoretical predictions and our hybrid simulation results, thereby validating the underlying physics of the scheme. These results open a new avenue for controlled laboratory studies of this important instability. 
Moreover, the demonstrated methodology offers an elegant, direct means to validate fundamental theories of parametric instabilities~\cite{Nishikawa1968}, which has wide-reaching relevance across plasma physics~\cite{hurricane2023physics,montgomery2016two,klimo2013laser,hasegawa1976parametric,sharma2021kinetic,clod2024cascades,chandran2018parametric,shoda2018frequency}, nonlinear optics~\cite{chirkin2000consecutive,wright2022physics,mckenna2022ultra,ledezma2022intense}, and nonlinear science~\cite{kadri2013generation,May_Qin_2024}.

\begin{acknowledgments}
This work was supported by DOE (grants DE-SC0023893, DE-SC0025443) and NASA (grant 80NSSC23K0695). 
X.F. acknowledges supports from NASA (grant 80NSSC23K0101) and NSF (grant 2229101). 
We acknowledge the Texas Advanced Computing Center (TACC) at The University of Texas at Austin 
and the National Energy Research Scientific Computing Center (NERSC) 
for providing the computing resources. 
\end{acknowledgments}


\appendix


\section{Effect of wave reflection on PDI measurement}\label{sec:reflection}

\begin{figure*}[htp]
	\centering
    \includegraphics[width=0.7\textwidth]{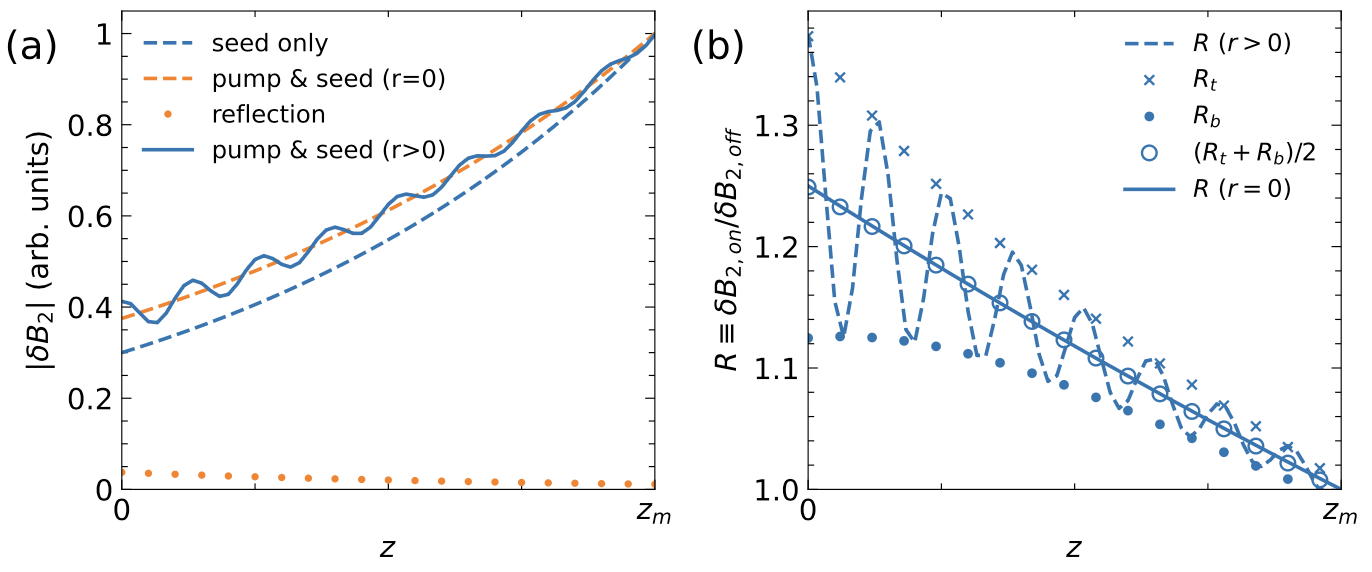}
	\caption{(a) Seed wave amplitudes vs $z$ under pump-off (blue dashed), pump-on \& without seed reflection (orange dashed), seed reflection only (orange dotted), and pump-on \& with reflection (blue solid) scenarios. (b) Corresponding seed amplitude ratios between ``pump-on \& with reflection'' and pump-off (dashed) runs, where the crosses, filled dots and open dots refer to the top, bottom envelopes of the modulation and their average, respectively. The blue solid curve stands for the seed wave ratios under ``pump-on \& without reflection''. See the main text for more details.}
	\label{fig4}
\end{figure*}

To illustrate how wave reflection affects the measurement of PDI, we construct a numerical model that allows for toggling the reflection on or off. 
As shown in Fig.~\ref{fig4}(a), we begin by injecting an initial seed wave at $z=z_m$, expressed as 
\begin{equation}
\label{seed-only}
    \delta B_2=\delta B_2(z_m)\exp[\gamma_d(z-z_m)]\cos(k_{\parallel2}z+\omega_2 t),
\end{equation}
where $\delta B_2(z_m)$ is the injection amplitude, $\gamma_d$ is the spatial damping rate, $z_m=30d_i$, $\omega_2/\Omega_{ci}\equiv\bar{\omega}_2=0.63$ and $k_{\parallel 2} d_i=\bar{\omega}_2/\sqrt{1-\bar{\omega}_2^2}$. 
At the left boundary $z=0$ (where the pump is injected and potential reflection may occur), the wave amplitude drops to $\delta B_2(z=0)=0.3\delta B_2(z_m)$. 
The seed wave amplitudes as a function of $z$ are represented by the blue dashed curve in the figure. 
In the presence of PDI but without reflection, the seed wave evolves into 
\begin{equation}
\label{seed_amplified_norefl}
    \delta B_2^\prime=\delta B_2(z_m)\exp[\gamma_d^\prime(z-z_m)]\cos(k_{\parallel2}z+\omega_2 t),
\end{equation}
where $\gamma_d^\prime<\gamma_d$ indicates reduced damping due to PDI growth. This results in a higher amplitude at $z=0$, specifically $\delta B_2^{\prime}(z=0)=0.375\delta B_2(z_m)$, depicted by the orange dashed curve.
Now, consider partial reflection of the seed wave at $z=0$ (orange dotted curve), modeled as
\begin{equation}
\label{reflected}
    \delta B_{2r}=r\delta B_2^{\prime}(z=0)\exp(-\gamma_dz)\cos(k_{\parallel2}z-\omega_2 t),
\end{equation}
with a reflection coefficient $r=0.1$. 
The superposition of the amplified seed wave and its reflection, $\delta B_2^\prime+\delta B_{2r}$, leads to a spatial modulation of the seed wave amplitude, shown as the blue solid curve in Fig.~\ref{fig4}(a).

This modulation is reflected in the seed amplitude ratios (pump-on vs pump-off), shown as the blue dashed curve in Fig.~\ref{fig4}(b). 
To isolate the PDI-induced amplification from this modulation, we apply an envelope averaging technique---calculating the mean of the top (crosses) and bottom (dots) of the oscillatory pattern. This average is plotted as open circles.
For validation, we also compare the seed wave ratios between a ``pump-on \& no-reflection'' case (orange dashed curve) and the baseline pump-off case (blue dashed curve) in Fig.~\ref{fig4}(a). The resulting ratio is plotted as the blue solid curve in Fig.~\ref{fig4}(b), which closely matches the open-circle curve. This agreement confirms that the envelope averaging procedure effectively removes reflection-induced artifacts in the measurement of PDI growth.

\bibliography{ref_pdi}

\end{document}